\documentclass[11pt,a4paper]{article}
\pdfoutput=1
\usepackage{jheppub}
\usepackage{graphicx} 
\usepackage{epstopdf} 
\usepackage{dcolumn} 
\usepackage{xcolor} 
\usepackage{amsmath,amssymb} 
\usepackage{hyperref} 
\usepackage[utf8]{inputenc} 
\usepackage[english]{babel} 
\usepackage[mathlines]{lineno} 
\usepackage{blindtext} 
\usepackage{ifthen} 
\usepackage{orcidlink} 
\usepackage[separate-uncertainty=true,multi-part-units=single]{siunitx}

\graphicspath{{figures/}} 

\newboolean{articletitles}
\setboolean{articletitles}{true} 

\input{belle2-symbols}

\newcommand{\lumi}{362 \pm 2}
\newcommand{\sqrts}{10.58}
\newcommand{\NDataMu}{\num{4371737}\xspace}
\newcommand{\NDataE}{\num{4358376}\xspace}
\newcommand{\data}{data\xspace}
\newcommand{\tabspace}{0.5cm}

\newcommand\taue      {\ensuremath{\tau^- \to e^- \bar\nu_e \nu_\tau}\xspace}
\newcommand\taumu     {\ensuremath{\tau^- \to \mu^- \bar\nu_\mu \nu_\tau}\xspace}
\newcommand\Btaue     {\ensuremath{{\cal B}(\taue)}\xspace}
\newcommand\Btaumu    {\ensuremath{{\cal B}(\taumu)}\xspace}

\newcommand\Be        {\ensuremath{{\cal B}_{e}}\xspace}
\newcommand\Bmu       {\ensuremath{{\cal B}_{\mu}}\xspace}

\newcommand\eetautau  {\ensuremath{e^+e^-\to\tau^+\tau^-}\xspace}

\newcommand\Rmu       {\ensuremath{R_\mu}\xspace}
\newcommand\RmuFull   {\ensuremath{\Rmu =\frac{\Btaumu}{\Btaue}}\xspace}

\newcommand\gmuge     {\ensuremath{|g_\mu /g_e|_\tau}\xspace}

\newcommand{\syseID}{0.22}
\newcommand{\sysmuID}{0.05}
\newcommand{\sysEmisID}{0.12}
\newcommand{\sysMumisID}{0.19}

\newcommand{\sysLID}{0.32}
\newcommand{\systrigger}{0.10}
\newcommand{\sysShape}{0.06}
\newcommand{\systag}{0.05}
\newcommand{\sysFSR}{0.08}
\newcommand{\sysISR}{0.01}
\newcommand{\sysnorm}{0.07}
\newcommand{\systrackeff}{0.01}
\newcommand{\sysMCSize}{0.06}
\newcommand{\syslumi}{0.01}
\newcommand{\sysPiEff}{0.02}

\newcommand{\relTotalSysLID}{0.32}
\newcommand{\relTotalSysSim}{0.14}

\newcommand{\RelSysTotal}{0.37}
\newcommand{\central}{0.9675}
\newcommand{\Stat}{0.0007}
\newcommand{\SysTotal}{0.0036}

\newcommand\result{\ensuremath{\central \pm \Stat \pm \SysTotal}\xspace}

\begin{document}

\vspace*{-3\baselineskip}
\resizebox{!}{2cm}{\includegraphics{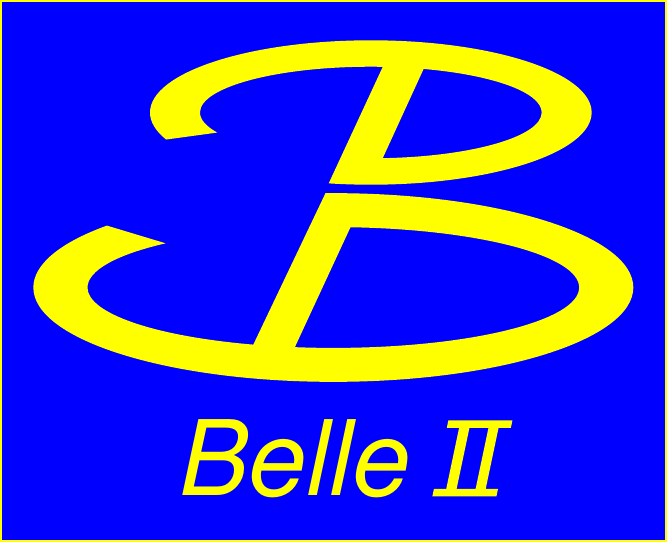}}


\title{Test of light-lepton universality in $\tau$ decays with the \belletwo experiment}
\collaboration{The Belle II Collaboration}
  \author{I.~Adachi\,\orcidlink{0000-0003-2287-0173},} 
  \author{K.~Adamczyk\,\orcidlink{0000-0001-6208-0876},} 
  \author{L.~Aggarwal\,\orcidlink{0000-0002-0909-7537},} 
  \author{H.~Aihara\,\orcidlink{0000-0002-1907-5964},} 
  \author{N.~Akopov\,\orcidlink{0000-0002-4425-2096},} 
  \author{A.~Aloisio\,\orcidlink{0000-0002-3883-6693},} 
  \author{N.~Anh~Ky\,\orcidlink{0000-0003-0471-197X},} 
  \author{D.~M.~Asner\,\orcidlink{0000-0002-1586-5790},} 
  \author{H.~Atmacan\,\orcidlink{0000-0003-2435-501X},} 
  \author{V.~Aushev\,\orcidlink{0000-0002-8588-5308},} 
% ...

\emailAdd{coll-publications@belle2.org}
\arxivnumber{2405.14625}

\abstract{%
We present a measurement of the ratio $\Rmu = \Btaumu / \Btaue$ of branching fractions $\cal B$ of the $\tau$  lepton decaying to muons or electrons using data collected with the \belletwo detector at the SuperKEKB \epem collider.
The sample has an integrated luminosity of \SI{\lumi}{\invfb} at a centre-of-mass energy of \SI{\sqrts}{\gev}.
Using an optimised event selection, a binned maximum likelihood fit is performed using the momentum spectra of the electron and muon candidates.
The result, $\Rmu = \result$, where the first uncertainty is statistical and the second is systematic, is the most precise to date.
It provides a stringent test of the light-lepton universality, translating to a ratio of the couplings of the muon and electron to the $W$ boson in $\tau$ decays of $0.9974 \pm 0.0019$, in agreement with the standard model expectation of unity.
}

\subheader{%
\vspace{-3cm}
\begin{flushright}
\normalfont
\belletwo Preprint 2024-002\\
KEK Preprint 2023-49
\end{flushright}
}

\maketitle
\flushbottom
\vspace{.05cm}

\section{Introduction}
\label{sec:into}

In the standard model (SM) of particle physics, lepton flavour universality (LFU) refers to an intrinsic property under which the electroweak gauge bosons have the same couplings to the three generations of leptons $e$, $\mu$, and $\tau$~\cite{Tsai:1971vv}.
This symmetry originates from the fact that the only difference between lepton generations is derived from their distinct masses.
A broad class of SM extensions postulates the existence of new particles, such as mediators of new interactions, that couple to the three leptons differently, making searches of LFU violation compelling~\cite{Jung:2010ik, Crivellin:2020klg}.

In the last decades, experimental tests of LFU have been carried out using various processes, but no significant deviation from the SM expectation has been observed.
For example, tests in charged currents have been carried out in the decays of pions~\cite{PiENu:2015seu}, kaons~\cite{NA62:2012lny, KLOE:2009urs}, $B$ mesons~\cite{Belle:2019rba,LHCb:2023zxo},
on-shell $W$ bosons~\cite{ATLAS:2020xea, CMS:2022mhs}, and $\tau$ leptons~\cite{CLEO:1996oro,BaBar:2009lyd}.
Among these, the most precise LFU test is achieved in pion decays, followed by the $\tau$-lepton decays, which are sensitive not only to charged currents but also to non-SM contributions of weak neutral currents~\cite{Altmannshofer:2016brv, Bryman:2021teu}.
The LFU tests in $\tau$ decays rely on measurements of the $\tau$ mass, lifetime, and branching fractions of $\tau$ decays to lighter leptons or hadrons.
The $e$-$\mu$ universality is tested by comparing the measured rates of leptonic $\tau$ decays, whose branching fractions are denoted as $\Btaumu$ and $\Btaue$.
Charge-conjugate modes are implied throughout the paper.
The ratio \Rmu of the branching fractions,
\begin{equation}
 \RmuFull,
\end{equation}
in turn, constrains the ratio of the effective coupling strengths $g_e$ and $g_\mu$ of the electron and muon to the $W^{\pm}$,
\begin{eqnarray}
 \label{eq:gmuge}
 \Bigg|\frac{g_\mu}{g_e}\Bigg|_\tau =\sqrt{ R_\mu \frac{f(m_e^2/m_\tau^2)}{f(m_\mu^2/m_\tau^2)}}.
\end{eqnarray}
Here $f(x)=1-8x+8x^3-x^4-12x^2 \ln{x}$ under the assumption of negligible neutrino masses~\cite{Tsai:1971vv}, and $m_e$, $m_\mu$, and $m_\tau$ are the masses of the corresponding leptons.
Thus, the challenge to test the $e$-$\mu$ universality is to accurately determine the ratio of branching fractions \Rmu.
This deviates from unity in the SM due to the difference in mass of the final state leptons and is predicted to be \num{0.9726}.
Previous measurements of \Rmu were reported by the CLEO~\cite{CLEO:1996oro} and BaBar~\cite{BaBar:2009lyd} collaborations.
The most precise single determination of \Rmu to date comes from the direct measurement of \Rmu by the BaBar collaboration, $\Rmu = 0.9796 \pm 0.0016 \pm 0.0036$~\cite{BaBar:2009lyd}, where the first uncertainty is statistical and the second is systematic.
The measurement has \SI{0.4}{\percent} precision limited by the systematic uncertainties associated mainly with the lepton identification.
This uncertainty propagates to \SI{0.2}{\percent} precision on $\gmuge=1.0036\pm 0.0020$.
The current world average value of $\gmuge = 1.0019 \pm 0.0014$~\cite{HeavyFlavorAveragingGroup:2022wzx} is consistent with unity, as predicted by the SM.

In this paper, we report a measurement of \Rmu using data collected with the \belletwo detector~\cite{Abe:2010gxa} at the energy-asymmetric \epem SuperKEKB collider~\cite{Akai:2018mbz}, with \SI{7}{\gev} electron and \SI{4}{\gev} positron beams colliding at \SI{4.76}{\degree}.
The data were recorded between 2019 and 2022 at a centre-of-mass energy of \SI{\sqrts}{\gev} and correspond to an integrated luminosity of \SI{\lumi}{\invfb} which translates to about \num{333e6} \eetautau events.
We determine \Rmu from events in which one $\tau$ decays either to \taue or \taumu, and the other $\tau$ decays hadronically.
First, we optimise a selection that is common to both modes, after which we perform a binned maximum likelihood fit using the momentum spectra of the lepton candidates.
All systematic effects are modelled directly in the likelihood function.
Finally, we measure the value of \Rmu from the fit and translate it into \gmuge, testing $e$-$\mu$ universality in charged current interactions.

\section{The Belle II detector and simulation}
\label{sec:BelleII}

The Belle II detector comprises several subdetectors arranged in a cylindrical structure around the $e^+e^-$ interaction point~\cite{Abe:2010gxa}.
Charged-particle trajectories (tracks) are reconstructed by a two-layer silicon-pixel detector, surrounded by a four-layer double-sided silicon-strip detector and a central drift chamber (CDC).
Only \SI{15}{\percent} of the second pixel layer was installed when the data were collected.
Outside the CDC, a time-of-propagation detector and an aerogel ring-imaging Cherenkov detector cover the barrel and forward endcap regions.
The electromagnetic calorimeter (ECL), divided into the forward endcap, barrel, and backward endcap, fills the remaining volume inside a \SI{1.5}{\tesla} superconducting solenoid and is used to reconstruct photons and electrons.
A $K_L^0$ and muon detection system is installed in the iron flux return of the solenoid.
The $z$ axis of the laboratory frame is defined as the detector solenoid axis, with the positive direction along the electron beam.
The polar angle $\theta$ and the transverse plane are defined relative to this axis.

Simulated samples are used for studying sample composition and optimising the analysis selections.
We further rely on simulated samples to study and determine efficiencies and to define fit templates for the extraction of \Rmu.
Several processes contribute to the \eetautau sample as backgrounds, including
$\epem \to q\bar{q}$ events, where $q$ indicates a $u$, $d$, $s$, or $c$ quark;
$\epem\to \epem (\gamma)$ and $\mu^+\mu^-(\gamma)$ events;
$\epem \to l^+l^-l^+l^-$ events, where $l$ is a charged lepton;
$\epem \to e^+e^-h^{+}h^{-}$ events, where $h$ indicates a pion or kaon; and
$\epem \to e^+e^- Nh$ events with multiplicity $N>2$.
We use several software packages to generate the simulated particles.
The \eetautau process is generated using KKMC~\cite{Jadach:1999vf, Jadach:2000ir}, $\tau$ decays are simulated by TAUOLA~\cite{Banerjee:2021rtn, Shekhovtsova:2012ra, Chrzaszcz:2016fte, Nugent:2013hxa} and their radiative corrections by PHOTOS~\cite{Barberio:1990ms}.
We use KKMC to simulate $\mu^+\mu^-(\gamma)$ and $q\bar{q}$ production; PYTHIA~\cite{Sjostrand:2014zea} for the fragmentation of the $q\bar{q}$ pair;
BabaYaga@NLO~\cite{Balossini:2006wc, Balossini:2008xr, CarloniCalame:2003yt, CarloniCalame:2001ny,CarloniCalame:2000pz} for $\epem \to \epem (\gamma)$ events;
and AAFH~\cite{BERENDS1985421,BERENDS1985441,BERENDS1986285} and TREPS~\cite{Uehara:1996bgt} for the production of non-radiative final states $l^+l^-l^+l^-$ and $e^+e^-h^{+}h^{-}$.
Currently, there is no generator to simulate $\epem \to e^+e^- Nh$ processes.
The \belletwo analysis software~\cite{Kuhr:2018lps, basf2-zenodo} uses the GEANT4~\cite{Agostinelli:2002hh} package to simulate the response of the detector to the passage of the particles.

\section{Event selection}
\label{sec:selection}

The trigger is based on ECL energy deposits (clusters) and their topologies in the ECL.
The efficiency of the trigger system for this measurement is driven by the condition that the combined energy deposit of all ECL clusters exceeds \SI{1}{\gev}.
Events with two back-to-back clusters in the centre-of-mass system, one of which exceeds \SI{4.5}{\gev} and the other \SI{3}{\gev}, are vetoed by the trigger system to reject Bhabha events.

In the \epem centre-of-mass frame, the $\tau$ leptons from \eetautau are produced in opposite directions and with a significant boost.
Thus, the decay products of one $\tau$ are isolated from those of the accompanying $\tau$ and contained in opposite hemispheres.
The boundary between those hemispheres is the plane perpendicular to the $\tau$ flight direction, which is experimentally approximated by the thrust axis. The thrust axis is the unit vector $\hat{t}$ that maximizes the thrust value $\sum{|\hat{t}\cdot \vec{p}_i^{\, *}|}/\sum{|\vec{p}_i^{\, *}|}$, where $\vec{p}_i^{\, *}$ is the momentum of the $i$th final state particle in the \epem centre-of-mass frame~\cite{Brandt:1964sa,Farhi:1977sg}.
This calculation uses charged particles with the requirements given later in this section, as well as photons identified from clusters with energy above \SI{150}{\mev} and within the CDC acceptance, having a polar angle $\theta$ of the momentum vector in the laboratory frame within $\SI{17}{\degree} < \theta < \SI{150}{\degree}$ to ensure they are not matched to any charged particle.
Throughout this paper, quantities in the \epem centre-of-mass frame are indicated by an asterisk.

We define the \emph{signal} hemisphere as the one containing a charged particle originating either from \taue or \taumu decays. We also require that the opposite hemisphere, labelled with \emph{tag}, contains only one charged particle and at least one neutral pion.
Thus, the tag side contains predominantly $\tau^+\to h^+ n \pi^0 \bar\nu_\tau$ decays with multiplicity $n= 1, 2$.

We select $\tau$-pair candidates by requiring the event to contain exactly two charged particles with zero total charge, each having a trajectory displaced from the average interaction point by less than \SI{3}{\cm} along the $z$ axis and less than \SI{1}{\cm} in the transverse plane to reduce the contribution of misreconstructed or poorly constrained tracks.
The charged particle on the tag side must satisfy the condition $E_{\rm ECL}/p \leq 0.8\,c$ to suppress electron contamination. Here, $E_{\rm ECL}$ denotes the energy deposit in the ECL, and $p$ is the magnitude of the momentum vector of the associated particle.
The charged particle on the signal side must be identified as either a muon or an electron according to the following conditions.
Muons are identified using the discriminator $P_\mu = {\cal L}_\mu / ({\cal L}_e + {\cal L}_\mu + {\cal L}_\pi + {\cal L}_K + {\cal L}_p + {\cal L}_d)$ where the likelihood ${\cal L}$ for each charged-particle hypothesis combines particle identification information from all detectors except the silicon trackers.
Electrons are identified using a boosted decision tree classifier that is trained to separate electrons from all other charged particles~\cite{eIDbdt}.
This approach gives improved results compared to the purely likelihood-based approach, in particular for separating pions from electrons.
Inputs to the classifier are the likelihoods from each sub-detector, as well as additional ECL observables, such as variables that characterise the cluster's spatial structure.
The most discriminating variable in the momentum range relevant to this analysis is $E_{\rm ECL}/p$.
We use the output of the classifier, $P_e$, as a discriminator for electron identification.
We retain lepton candidates with requirements $P_e > 0.5$ and $P_\mu > 0.9$. We require that each lepton candidate, $\ell = e, \mu$ in the signal side satisfies the conditions $1.5< p_\ell < \SI{5}{\gevc}$, and $0.82 < \theta_\ell < \SI{2.13}{\rad}$ to ensure accurate particle identification information.
Here $\theta_\ell$ refers to the polar angle of the momentum vector in the laboratory frame.
While inefficiencies of the particle identification system are taken into account in our simulation, we correct for imperfections in the simulation using data-driven factors from calibration channels, as functions of momentum, polar angle, and charge.
These channels are $J/\psi\rightarrow \ell^+\ell^-$, $\epem \to \ell^+\ell^-\gamma$, and $\epem \to \epem\ell^+\ell^-$ events for efficiency, and $K^{0}_{S} \rightarrow \pi^{+}\pi^{-}$ and $\eetautau$ events for misidentification rates.
Due to limitations in calibration-sample sizes, events with candidate leptons $p_\ell > \SI{4}{\gevc}$ and $\theta_\ell > \SI{1.78}{\rad}$, or $p_\ell > \SI{4.5}{\gevc}$ and $\theta_\ell > \SI{1.16}{\rad}$ are vetoed.
Electron and muon identification efficiencies are \SI{99.7}{\percent} and \SI{93.9}{\percent}, respectively. The rates for misidentifying
pions as electrons or muons are \SI{0.9}{\percent} and \SI{3.1}{\percent}, respectively.

The momenta of charged particles are corrected for imperfections in the magnetic field description used for event reconstruction, misalignment of the detector, and material mismodelling. The corresponding correction factor is evaluated by measuring the mass-peak position of a high-yield sample of $D^{0}\rightarrow K^{-} \pi^{+}$ decays reconstructed in data and comparing this to the known value~\cite{ParticleDataGroup:2022pth}.

Neutral pions are identified as photon pairs with an invariant mass between \SI{120}{\mevcc} and \SI{145}{\mevcc}, which is within two units of mass resolution from the known value.
Those photons are identified from clusters, reconstructed within the CDC acceptance, $17^\circ < \theta < 150^\circ$, to ensure they are not matched to any charged particle.
The energy threshold for selecting photon candidates varies based on the polar detector region, aiming to suppress beam-induced backgrounds. This adjustment is important for the endcaps where such backgrounds are more prominent. Specifically, we use \SI{80}{\mev} for the forward region with $\SI{17}{\degree} < \theta < \SI{31.4}{\degree}$, \SI{30}{\mev} for the barrel with $\SI{32.2}{\degree} < \theta < \SI{128.7}{\degree}$, and \SI{60}{\mev} for the backward region with $\SI{130.7}{\degree} < \theta < \SI{150}{\degree}$.
The requirements $\alpha_{\gamma\gamma} < \SI{1.4}{\rad}$ and $|\delta \phi| < \SI{1.5}{\rad}$ on the angle $\alpha$ between the momenta of the two photons, and on the difference of azimuthal angles $\phi$ of the two photons reduce the combinatorial background from low-energy photons.
The efficiency for identifying neutral pions with these criteria is \SI{30}{\percent}.

The main background contamination comes from $\epem\to \epem (\gamma)$, $\epem \to e^+e^- l^+ l^-$ and $\epem \to \epem Nh$ processes.
The $\epem \to \epem Nh$ events are suppressed using a data-driven selection.
For these processes, the events are concentrated at low thrust values, at large values of the squared missing mass $M^2_{\rm miss}$, and at low missing transverse momentum $p^*_{T, \; \rm miss}$.
The missing momentum is the difference between the momenta of the initial \epem and that of all reconstructed particles in the event, while the square of the missing mass is defined as $M^2_{\rm miss} = (\sqrt{s}/c^2-E^*_{\rm vis}/c^2)^2-(p^*_{\rm miss}/c)^2$, where $E^*_{\rm vis}$ is the energy of all reconstructed particles in the event and $p^*_{\rm miss}$ is the magnitude of the missing momentum vector.
The $\epem \to \epem Nh$ background can be discriminated with these variables since the \epem pair is not reconstructed and has a momentum vector parallel to the beam axis, while the remaining particles are not collinear.
The requirements that the thrust value exceeds $0.85$ and $M^2_{\rm miss} < (\SI{20}{\gevgevcccc} + 40\,\gev/c^3 \cdot p^*_{T,\,\rm{miss}})$ are chosen to remove the contribution from the $\epem \to \epem Nh$ events.

\begin{figure}
 \centering
 \includegraphics[width=.768\linewidth]{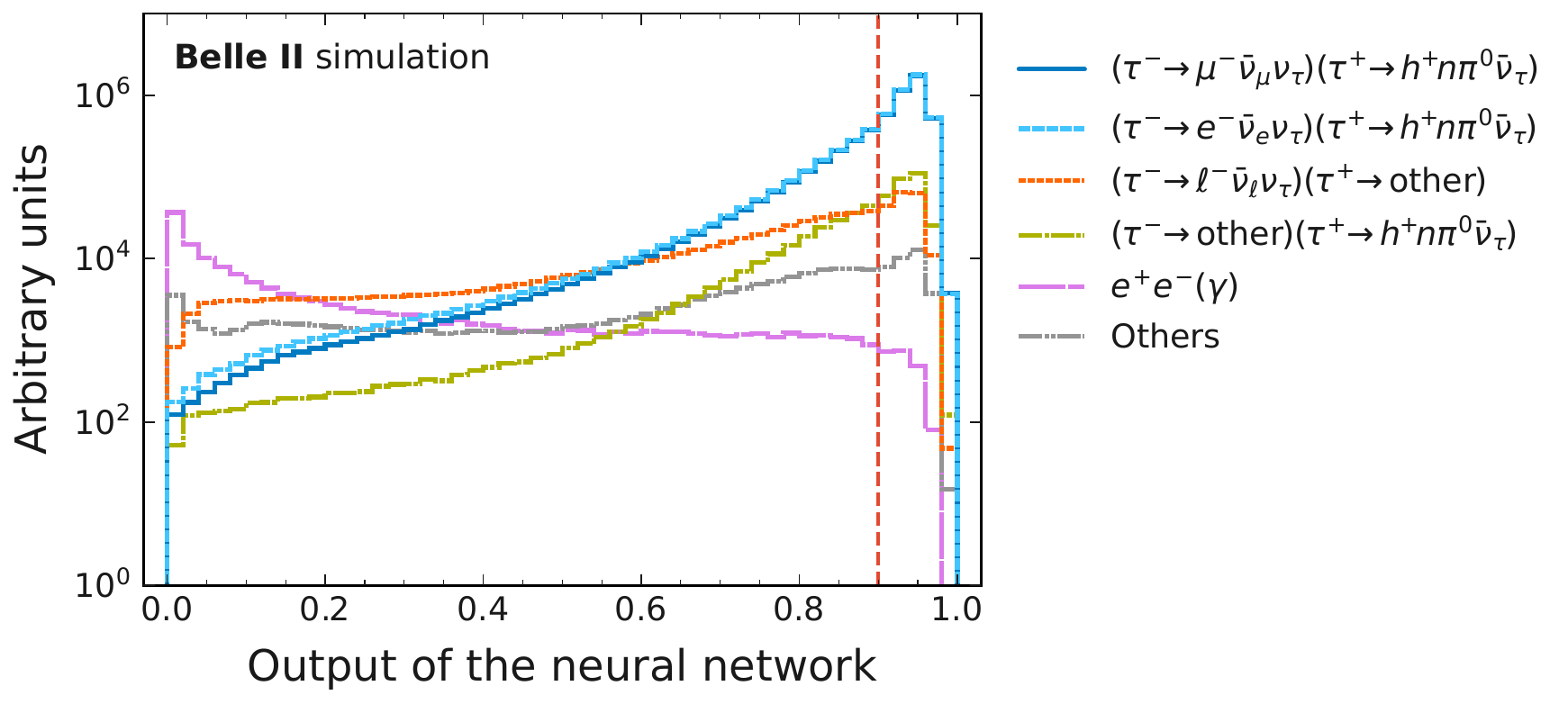}
 \caption{Distributions of the output of the neural network, trained to classify the combined sample of \taue and \taumu decays. We show simulated background contributions from $e^+e^- \to \tau^+\tau^-$ events with decays other than $\tau^+\to h^+ n \pi^0 \bar\nu_\tau$ on the tag side, with decays other than \taue and \taumu on the signal side, and Bhabha events separately.
  The remaining background processes contributing to the spectrum are combined and collectively called `Others'.
  The dashed vertical line indicates the threshold used in this analysis.}
 \label{fig:nn_outputs}
\end{figure}

For all other background processes, simulated signal and background events are used to train a neural network event classifier using cross-entropy as a loss function~\cite{Good:logloss, DBLP}.
The signal sample is defined as the combination of the \taue and \taumu samples to have a common selection for both decays and, as a result, a cancellation of most of the systematic uncertainties on \Rmu that are associated with the selection.
Seven variables are used in the training: the thrust value, the polar angle of the thrust vector, $E^*_{\rm {vis}}$, the transverse component of the missing momentum direction in the centre-of-mass frame, the momentum of the tag side charged particle and the invariant mass and polar angle of the $h^+ n \pi^0$ system on the tag side in the centre-of-mass frame.
We map them to four output nodes --- one representing the combined electron and muon signal sample, the other three representing the various background events: \eetautau events with a misidentified signal or tag side, $\epem \to \epem$ events, and a combined sample of all remaining background events.
The most discriminating variable is $E^*_{\rm {vis}}$, which is twice as important as the least discriminating variable.
Consistency checks with validation samples show no indication of overtraining.
Figure~\ref{fig:nn_outputs} shows the output distribution of the neural network used to discriminate between \eetautau events with \taue or \taumu decays on the signal side and all other events.
The figure uses simulated training samples.

\begin{figure}
 \centering
 \includegraphics[width=0.91\linewidth]{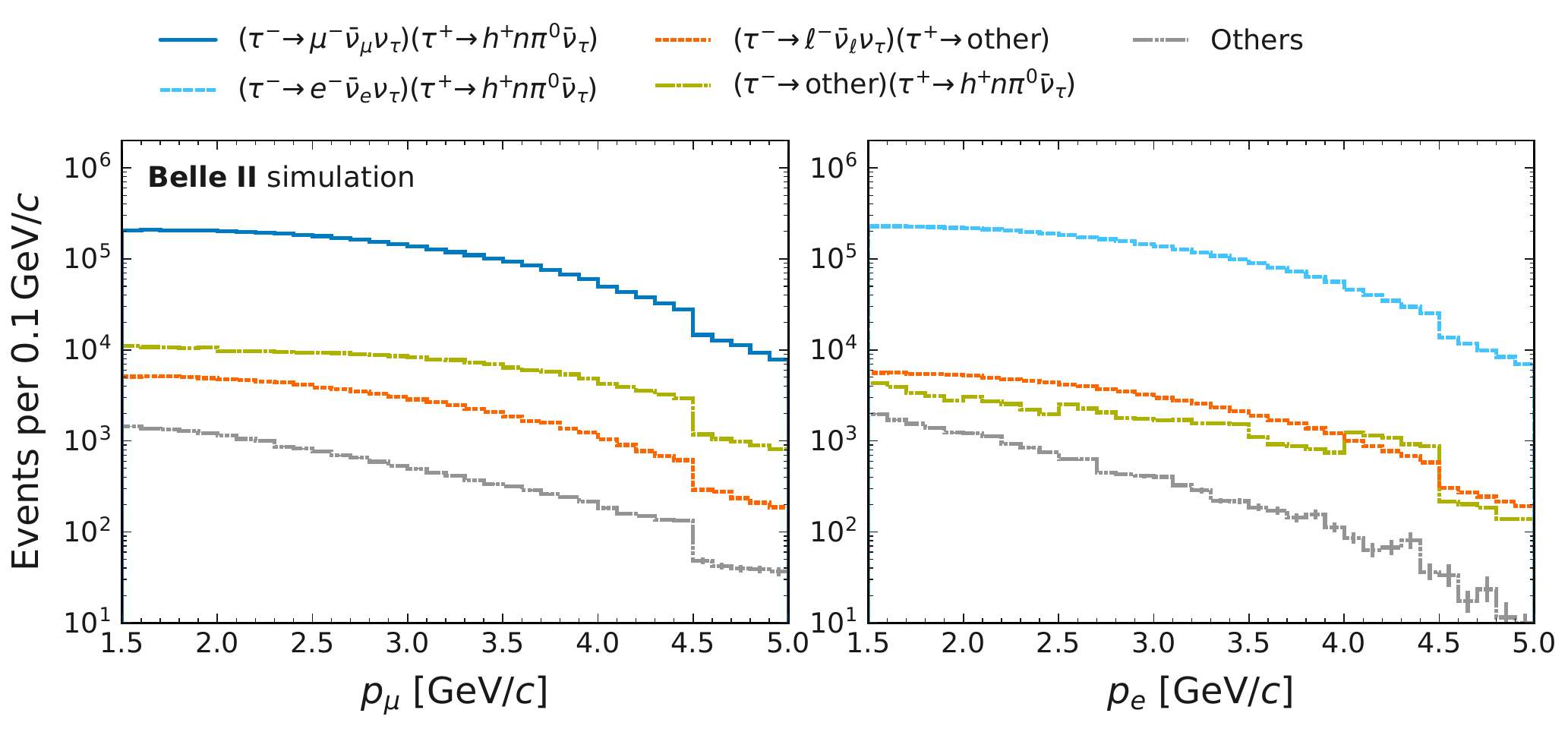}
 \caption{Momentum spectra of muon (left) and electron (right) candidates from simulated \taumu and \taue samples, along with simulated background contributions.
  The legend uses a similar nomenclature as Figure~\ref{fig:nn_outputs}.
 }
 \label{fig:dataMC_Pcombined}
\end{figure}

To select signal events, we require the output value of the neural network to be greater than 0.9, which yields the smallest total uncertainty on \Rmu in simulation.
The selected events are separated into electron and muon samples.
These samples will be simultaneously fit in bins of the lepton candidate momentum in order to extract \Rmu from the data.
The momentum distributions of the electron and muon candidates in the simulated \taue and \taumu samples, along with simulated background contributions, are shown in Figure~\ref{fig:dataMC_Pcombined}.
The distributions are corrected for imperfections in the simulation, in particular particle identification, the reconstruction of neutral particles, and trigger.
The discontinuities at \SI{4}{\gevc} and \SI{4.5}{\gevc} in the distributions reflect the veto of events in certain lepton identification correction bins as well as the impact of the correction factors themselves.

The trigger efficiency is measured with a reference sample selected by independent triggers based on the number of particles reconstructed in the CDC.
The trigger efficiency in \data is \SI{99.8}{\percent} for \taue and \SI{96.6}{\percent} for  \taumu decays,
which is primarily driven by the tag hemisphere.
In simulation, the corresponding efficiencies are \SI{98.6}{\percent} and \SI{95.4}{\percent}, respectively.
To account for imperfection in the simulation of the trigger, we apply correction factors as ratios of efficiencies in data and simulation to our simulated samples.
The correction factors are stable as a function of the lepton candidate momentum and independent of the flavour of the signal lepton.

The signal-reconstruction efficiency in simulation after all selection requirements and applied corrections is \SI{9.602}{\percent} for \taue and \SI{9.551}{\percent} for \taumu.
The purity is \SI{96}{\percent} for the $e$-sample and \SI{92}{\percent} for the $\mu$-sample.
The dominant backgrounds are from $e^+e^- \to \tau^+\tau^- $ events with $\tau^-\to h^- \nu_\tau$ or $\tau^-\to h^- n \pi^0 \nu_\tau$ decays in the signal hemisphere, which makes up \SI{1.3}{\percent} of the $e$-sample and \SI{5.2}{\percent} of the $\mu$-sample.
This is followed by \SI{2.3}{\percent} of $e^+e^- \to \tau^+\tau^- $ events with a misidentified tag side decay and \SI{0.2}{\percent} of $e^+e^- \to e^+e^- \tau^+\tau^-$  events, for both samples. In \data, we observe \NDataMu events for which the lepton candidate is identified as a muon and \NDataE events for which it is identified as an electron.

\section{Method}
\label{sec:Rmu}

\begin{figure}
 \centering
 \includegraphics[width=.48\linewidth]{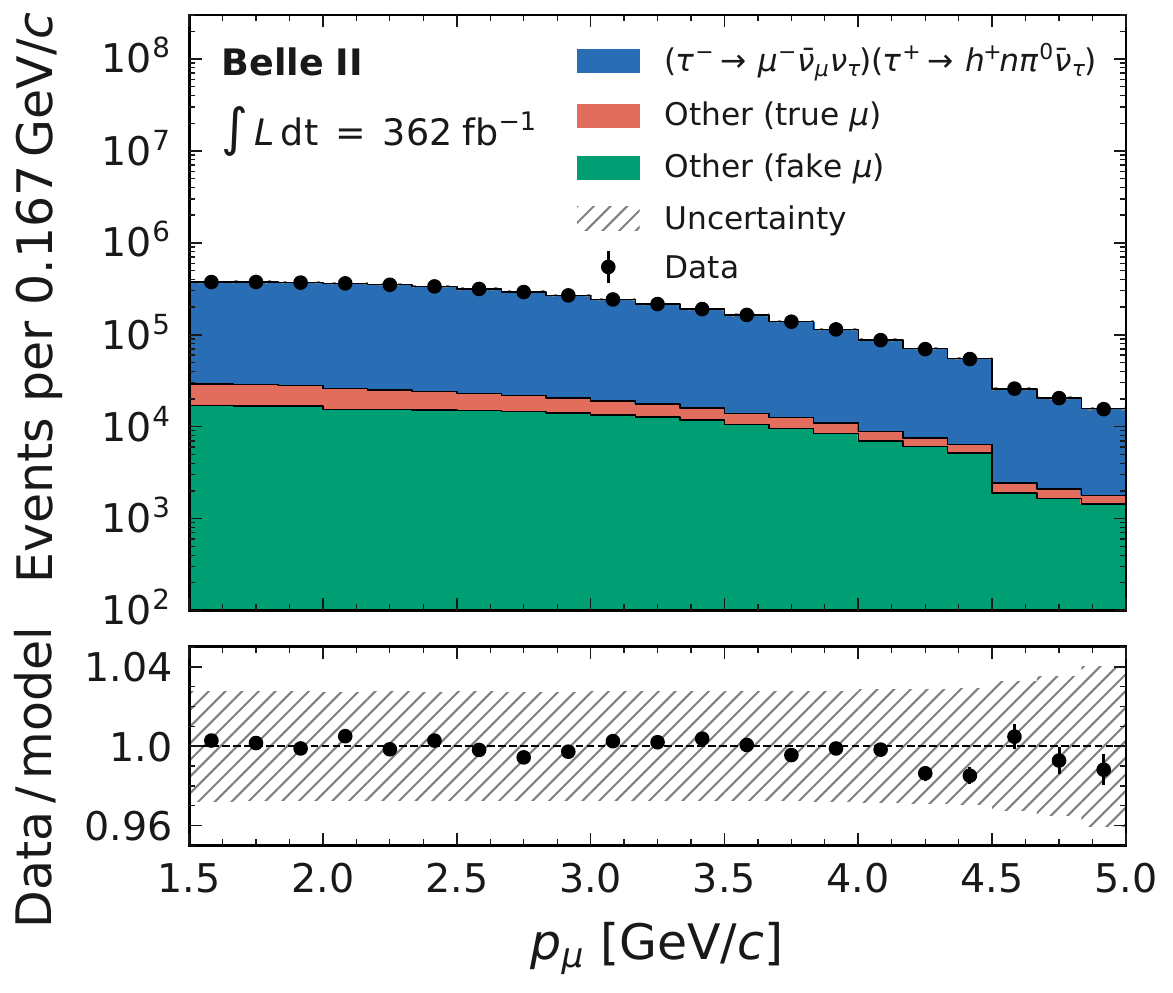}\hfill
 \includegraphics[width=.48\linewidth]{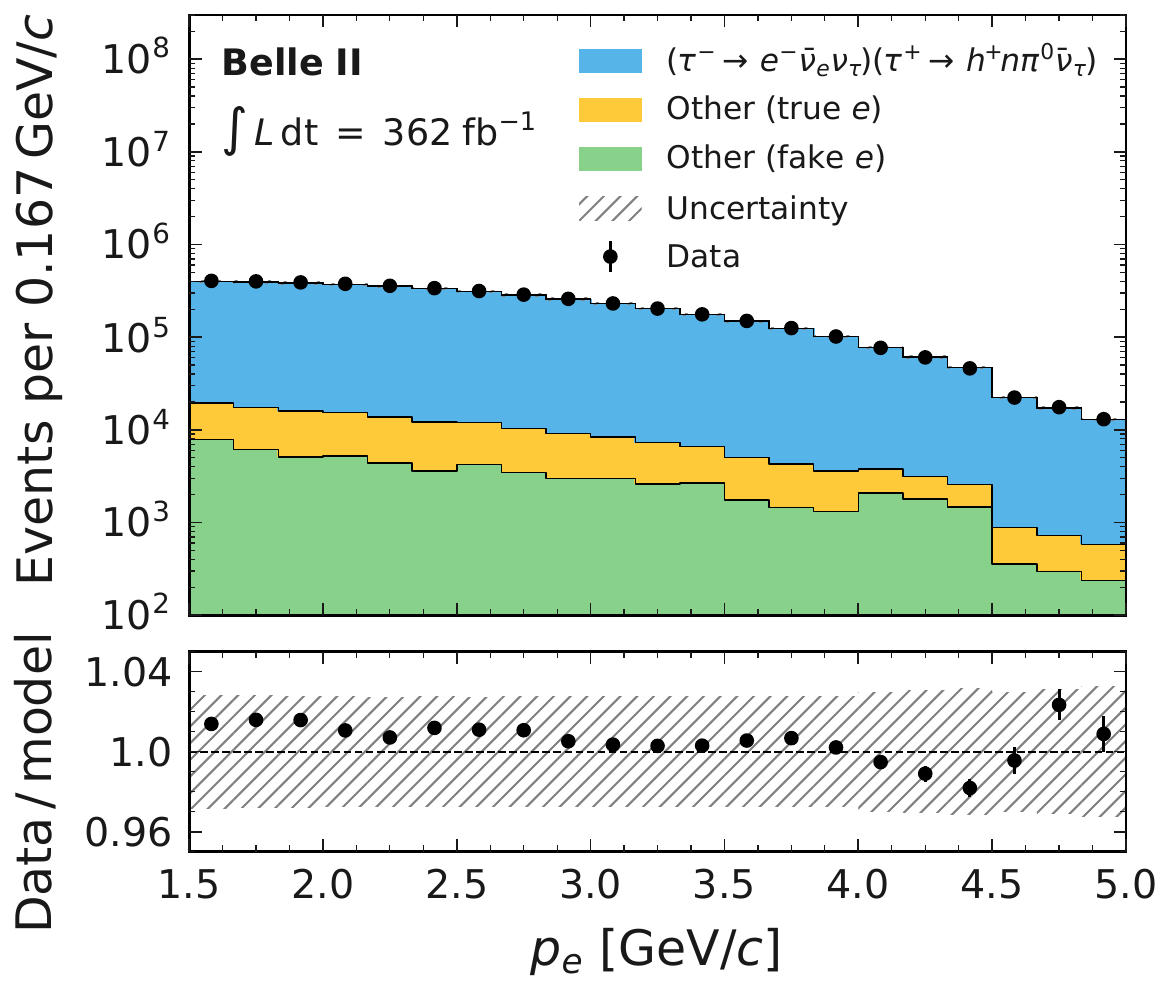}
 \caption{Observed momentum distribution for muon (left) and electron (right) candidates with simulation expectations overlaid.
  The lower panel shows the ratio between data and expectations with systematic uncertainties (hatched).
 }
 \label{fig:dataMC_prefit}
\end{figure}

To determine signal yields we use the \emph{pyhf} package~\cite{Heinrich:2021gyp}, which constructs a binned likelihood following the HistFactory~\cite{Cranmer:2012sba} formalism.
The templates for the signal and background momentum distributions are derived from simulation and use \num{21} bins spanning the lepton candidate momentum.
We apply data-driven corrections to these templates that account for imperfect simulation of particle identification, neutral particle reconstruction, and trigger efficiency.
This binning choice is derived from the lepton identification correction bins.
The momentum range is covered by seven bins with a bin width of \SI{0.5}{\gevc}, which are further split into three equal-sized bins to define the templates.
The events are separated into channels according to the previously specified lepton identification criteria.
For each channel, one template represents the signal component, and two templates describe the backgrounds, separated into events with correctly identified leptons and all other background events.
Systematic uncertainties are incorporated with multiplicative or additive event-count modifiers in the likelihood.
The likelihood function $f$ is a product of Poisson probability density functions $\cal P$ that combines the information from all bins of signal and background samples,
\begin{equation}
 f(\Rmu,\vec{\chi}) =
 \prod_{b\in {\rm bins}}{\cal P}(n_b^e   \mid \nu_b^e(\vec{\chi}))
 \prod_{b\in {\rm bins}}{\cal P}(n_b^\mu \mid \nu_b^\mu(\Rmu,\vec{\chi}))
 \prod_{\chi\in\vec{\chi}}c_\chi(a_\chi \mid \chi) .
 \label{eq:likelihood}
\end{equation}
Here, $n_b^e$ and $n_b^\mu$ are the number of observed \taue and \taumu candidate events in \data for each bin $b$ of the template. The corresponding expected number of events from simulations are denoted as $\nu_{b}^e$ and $\nu_{b}^\mu$. The systematic uncertainties discussed below are included in the likelihood as a set of nuisance parameters $\vec\chi$ that are event-count modifiers, constrained by normal probability density functions $c_\chi$ associated with auxiliary data $a_\chi$.
We model each momentum spectrum as a sum of contributions from the signal \taue or \taumu decays and all sources of background events.
Since the majority of the background originates from \eetautau events, we do not split the background templates according to the physics processes.
Instead, we distinguish them by the signal side particle type in order to separate the effects of the uncertainties associated with lepton identification and misidentification on the templates.
The resulting expressions for the expected events for each bin are
\begin{eqnarray}
 \nu_b^e(\vec{\chi}) &=&
 \kappa_e \, \nu_b^{e\text{-sig}}(\vec{\chi})
 + \nu_b^{e\text{-bkg(true)}}(\vec{\chi})
 + \nu_{b}^{e\text{-bkg(fake)}}(\vec{\chi}) \, \text{and} \\
 \nu_b^\mu(\Rmu, \vec{\chi}) &=& \Rmu \, \kappa_{e/\mu}^{\rm gen} \,
 \kappa_e \, \nu_b^{\mu\text{-sig}}(\vec{\chi})
 + \nu_b^{\mu\text{-bkg(true)}}(\vec{\chi})
 + \nu_b^{\mu\text{-bkg(fake)}}(\vec{\chi})\;.
\end{eqnarray}
Here $\nu_b^{e\text{-sig}}$ and $\nu_b^{\mu\text{-sig}}$ are the signal yields,
$\nu_b^{e\text{-bkg(true)}}$ and $\nu_b^{\mu\text{-bkg(true)}}$ are the yields of background events with the particle in the signal side properly identified as either an electron or a muon,
and $\nu_{b}^{e\text{-bkg(fake)}}$ and $\nu_b^{\mu\text{-bkg(fake)}}$  stand for the remaining background events, mostly with pions misidentified as leptons.
The factor $\kappa_e$, which is free in the fit, sets the overall normalisation of the signal templates; the constant scaling factor $\kappa_{e/\mu}^{\rm gen} \equiv {{\cal B}^{\rm gen}_e} / {{\cal B}^{\rm gen}_\mu}$ takes the assumed branching fractions from the simulation into account.
This allows the ratio \Rmu to be estimated directly from the fit: it is determined simultaneously along with the nuisance parameters by maximising the likelihood function.
Figure~\ref{fig:dataMC_prefit} shows the yields from experimental data, superimposed with the expected yields from simulation.
The hatched area in the lower panel indicates the possible variation of the yields due to systematic effects, with the dominant contribution being the $\pi^0$ efficiency uncertainty.

To minimise the observer-expectancy effect, we validate the \Rmu measurement method using simulation, and estimate the statistical and systematic uncertainties using \data without examining the central value of the result. This is achieved by randomly adjusting the values of branching fractions $\Be^{\rm gen}$ and $\Bmu^{\rm gen}$ used in the generation within a \SI{2}{\percent} range.
This allows us to validate the fit and the shapes of all distributions without observing the actual value.

\section{Systematic uncertainties and consistency checks}
\label{sec:sys}

The systematic uncertainties are grouped into categories associated with charged-particle identification, imperfections of the simulation, trigger, size of the simulated samples, and luminosity.
We include one or more nuisance parameters for each systematic source depending on the correlations between the individual template bins.
The nuisance parameters modify the event-counts of the templates to capture the impact of each systematic uncertainty on the observed momentum distribution of the leptons.
The total number of nuisance parameters in the fit amounts to \num{194}.
Table~\ref{table:systsum} summarises the sources contributing to the total uncertainty on \Rmu.
To estimate the uncertainty of each contribution, we generate \num{2000} simulated replicas of data by sampling the likelihood, where only auxiliary data associated with the respective nuisance parameters are varied according to the associated constraint terms.
The standard deviation of the resultant distribution of fitted \Rmu values is quoted as the size of the corresponding systematic effect.
The largest uncertainty on \Rmu arises from the particle identification.
The total absolute systematic uncertainty on \Rmu is $\SysTotal$.

\begin{table}
 \centering
 \begin{tabular}{lr}
  \hline
  \vspace{-0.45 cm}                                                                               \\
  Source                                                    & Uncertainty [\%]                    \\[0.04cm]
  \hline
  \vspace{-0.45 cm}                                                                               \\
  Charged-particle identification:                          & \relTotalSysLID \hspace*{\tabspace} \\[0.04cm]
  \hspace*{\tabspace}Electron identification                & \syseID                             \\
  \hspace*{\tabspace}Muon misidentification                 & \sysMumisID                         \\
  \hspace*{\tabspace}Electron misidentification             & \sysEmisID                          \\
  \hspace*{\tabspace}Muon identification                    & \sysmuID                            \\[0.04cm]
  \hline
  \vspace{-0.45 cm}                                                                               \\
  Imperfections of the simulation:                          & \relTotalSysSim \hspace*{\tabspace} \\[0.04cm]
  \hspace*{\tabspace}Modelling of FSR                       & \sysFSR                             \\
  \hspace*{\tabspace}Normalisation of individual processes  & \sysnorm                            \\
  \hspace*{\tabspace}Modelling of the momentum distribution & \sysShape                           \\
  \hspace*{\tabspace}Tag side modelling                     & \systag                             \\
  \hspace*{\tabspace}$\pi^0$ efficiency                     & \sysPiEff                           \\
  \hspace*{\tabspace}Particle decay-in-flight               & 0.02                                \\
  \hspace*{\tabspace}Tracking efficiency                    & 0.01                                \\
  \hspace*{\tabspace}Modelling of ISR                       & \sysISR                             \\
  \hspace*{\tabspace}Photon efficiency                      & $< 0.01$                            \\
  \hspace*{\tabspace}Photon energy                          & $< 0.01$                            \\
  \hspace*{\tabspace}Detector misalignment                  & $< 0.01$                            \\
  \hspace*{\tabspace}Momentum correction                    & $< 0.01$                            \\[0.04cm]
  \hline
  \vspace{-0.45 cm}                                                                               \\
  Trigger                                                   & \systrigger \hspace*{\tabspace}     \\[0.04cm] \hline
  \vspace{-0.45 cm}                                                                               \\
  Size of the simulated samples                             & \sysMCSize \hspace*{\tabspace}      \\[0.04cm] \hline
  \vspace{-0.45 cm}                                                                               \\
  Luminosity                                                & \syslumi \hspace*{\tabspace}        \\[0.04cm] \hline
  \vspace{-0.45 cm}                                                                               \\
  Total                                                     & \RelSysTotal \hspace*{\tabspace}    \\[0.04cm] \hline
 \end{tabular}
 \caption{Fractional systematic uncertainty on \Rmu, split into the contributing sources.
  The total fractional systematic uncertainty is \SI{\RelSysTotal}{\percent}, translating into an absolute uncertainty of $\SysTotal$.}
 \label{table:systsum}
\end{table}

\subsection{Charged-particle identification}

\begin{figure}
 \centering
 \includegraphics[width=.5\linewidth]{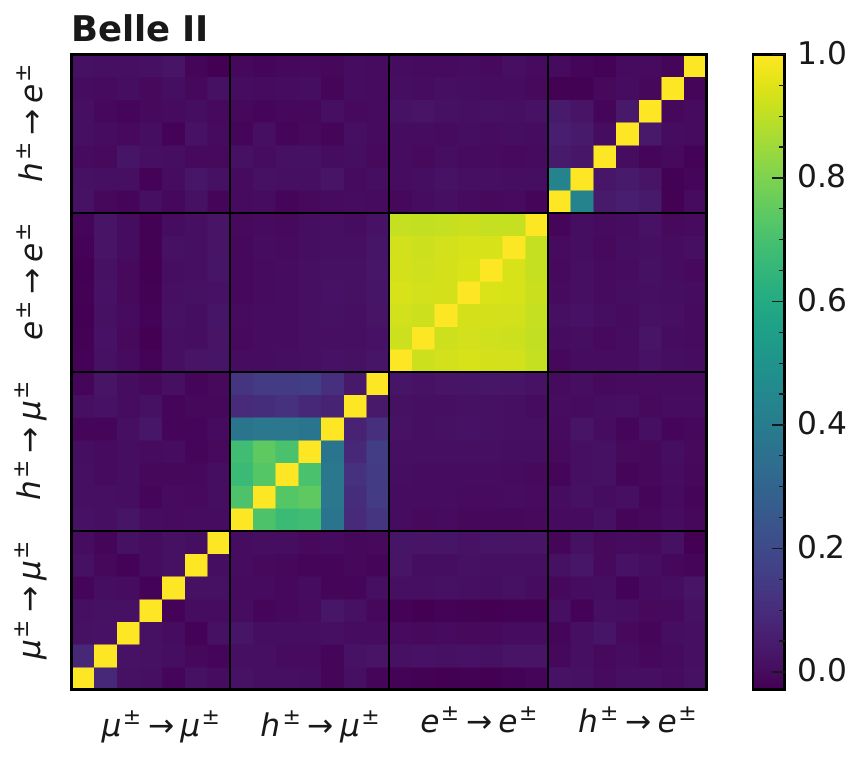}
 \caption{Correlations between the correction uncertainties of the lepton efficiencies ($\mu^{\pm} \to \mu^{\pm}$, $e^{\pm} \to e^{\pm}$)  and misidentification rates ($h^{\pm} \to \mu^{\pm}$, $h^{\pm} \to e^{\pm}$).
  Each block for every category shows the correlation between the individual momentum bins.
 }
 \label{fig:corr_matrix}
\end{figure}

Systematic uncertainties associated with charged-particle identification are obtained from data-driven corrections to the lepton identification efficiencies and particle misidentification rates in simulation.
The corrections and their uncertainties are summarised in Table~\ref{table:lepid}.
The uncertainties of the correction factors have statistical and systematic components.
The statistical uncertainties are dominant for the muon identification corrections and are included in the fit model as fully independent across all correction bins.
Systematic components dominate the uncertainties of the electron identification corrections.
The systematic uncertainties are assumed to be fully correlated across the correction bins, which is demonstrated in simplified simulated experiments to be the most conservative approach.
If the uncertainty is asymmetric, the larger of the signed variations is considered for both.
The resulting correlations between the individual template bins are shown in Figure~\ref{fig:corr_matrix}.
The efficiency correction uncertainties are applied to the templates with true leptons of the same type.
Meanwhile, misidentification rate uncertainties are applied to the templates with the fake leptons.
Since the corrections are binned in steps of \SI{0.5}{\gev} in momentum, only 7 parameters are required to model the uncertainties associated with efficiencies and misidentification rates, respectively.
This results in a total of 28 nuisance parameters.
The relative uncertainty on \Rmu associated with charged-particle identification is \SI{\sysLID}{\percent}, shared between the muon and electron identification and misidentification uncertainties.

\begin{table}[thb]
 \renewcommand{\arraystretch}{1.05}
 \centering
 \begin{tabular}{lcccc}
  \hline
                             & Correction             & Average    & Uncertainty            & Average          \\
                             & range                  & correction & range [\%]             & uncertainty [\%] \\
  \hline
  Electron identification    & \SIrange{0.99}{1.01}{} & \num{1.00} & \SIrange{0.05}{1.35}{} & \num{0.22}       \\
  Muon identification        & \SIrange{0.87}{0.99}{} & \num{0.97} & \SIrange{0.10}{6.00}{} & \num{0.38}       \\
  \hline
  Electron misidentification & \SIrange{1.00}{13.0}{} & \num{2.61} & \SIrange{3.54}{162}{}  & \num{41.4}       \\
  Muon misidentification     & \SIrange{0.25}{1.10}{} & \num{0.82} & \SIrange{3.11}{76.9}{} & \num{11.7}       \\
  \hline
 \end{tabular}
 \caption{Ranges and average values for the correction factors and associated total uncertainties of the lepton identification efficiencies, and the rates of pions misidentified as leptons. The uncertainties are given relative to the correction factors.}
 \label{table:lepid}
\end{table}

\subsection{Imperfections of the simulation}

Any mismodelling of the simulation that affects both the electron and the muon momentum distributions equally would cancel in measuring \Rmu.
We examine potential simulation mismodellings that could impact the momentum distributions of electrons and muons in \taue and \taumu decays or one of the backgrounds differently.
Specifically, modelling of initial and final state radiation (ISR and FSR), and the reconstruction efficiency of radiated photons may differ in simulated samples and \data.
We modify the number of events with an ISR photon by varying them from \SIrange{5}{20}{\percent}.
The resulting effect on \Rmu is approximately \SI{\sysISR}{\percent} for all variations, since ISR largely cancels because it has a similar effect on all templates.
This is not true for FSR, which also includes bremsstrahlung photons, since the effect on the momentum of the electron in \taue is expected to be larger than that of the muon in \taumu.
To account for the possibility of misestimating the number of FSR photons in the simulation compared to \data, we randomly select \SI{5}{\percent} of the generated FSR photons in simulation and add their momenta to the momentum of the corresponding lepton.
This fraction of added FSR momenta is based on a study with $J/\psi \to \epem$ and $J/\psi \to \mu^+ \mu^-$ samples, with the second as a control mode not sensitive to radiative effects in the final state. Agreement within \SI{5}{\percent} is observed between simulations and \data for the control samples in the distribution of the energy recovered as FSR from electrons.
The resulting template variations of the lepton momentum distributions are symmetrised to also account for the possible underestimation of the number of FSR photons in the simulation.
The systematic uncertainty arising from the potential mismodelling of FSR propagated to \Rmu is estimated to be \SI{\sysFSR}{\percent}.

Systematic effects may emerge due to the differences in the normalisations of the underlying physics processes.
Here, we independently consider variations in the yields of \eetautau events with a misidentified signal side, with a misidentified tag side, as well as the yields of non-taupair background processes. In addition, we vary the total \eetautau yield given the uncertainty of the \eetautau cross section and the uncertainties of the tag side branching fractions.
We assess the effect of these normalisation differences by recomputing the templates with the varied yields.
The resulting variations are included in the model, where each independent variation is parametrised by one nuisance parameter.
The corresponding systematic uncertainty on \Rmu is \SI{\sysnorm}{\percent}.

To account for possible simulation mismodelling that could alter the shape of the templates, we add uncertainties to each template bin that are fully correlated across all six templates but only affect one momentum bin at a time, resulting in \num{21} nuisance parameters.
The size of this uncertainty is estimated in a data-driven manner by analysing the spread of the ratio between \data and simulations of a sample without any particle identification requirement imposed on the signal side.
We use the standard deviation of this ratio, computed in the \num{21} momentum bins of the templates, which amounts to \SI{0.84}{\percent}.
This results in an overall effect on \Rmu of \SI{\sysShape}{\percent}.

Using an \emph{embedding} technique on the combined sample of \taue and \taumu events, both in simulations and in \data, we study possible simulation mismodellings of the tag side that might affect the event selection.
We select the electron or muon candidate in each event and replace it with a randomly selected lepton from the simulated \taue or \taumu events.
To ensure a reasonable match, we perform these replacements in quintiles of the lepton momentum, selecting charged particles with momentum magnitudes similar to the original lepton.
We then adjust the momentum direction of the replaced particle to align with the original momentum direction. Additionally, all variables affected by the replaced lepton momentum are recalculated accordingly.
The observed differences of the resulting data-over-simulation ratios between the embedded \taue and \taumu events are around \SI{0.2}{\percent}, depending on the momentum of the lepton candidate.
Since the same \taue or \taumu events are embedded in \data and simulated samples, any differences are expected to arise solely from the tag side.
These differences are
small and consistent with statistical fluctuations, but to be conservative they are
included in the fit model as fully uncorrelated variations of the template bins with \num{42} nuisance parameters and contribute a \SI{\systag}{\percent} uncertainty on \Rmu.

The correction of neutral pion reconstruction efficiency, to account for differences between simulation and \data, is determined as a function of the pion momentum from auxiliary measurements in calibration channels of $D$ and $\tau$ decays.
The impact of the uncertainty of this correction on \Rmu is estimated to be \SI{\sysPiEff}{\percent}.

To estimate the impact of mismodelled charged-particle decay-in-flight rates in simulations, we modify the number of tracks reconstructed from the decay products of other particles, primarily pions, by \SI{5}{\percent}. The mismodelling of tag side pion decays affects the muon and electron templates similarly, leading to a cancellation of the systematic effects. However, the signal side muon templates would be affected significantly by the modification in pion decay rates, and the effect does not cancel out. The associated systematic uncertainty on \Rmu is \SI{0.02}{\percent}.

Differences between the track finding efficiencies in simulation and \data have been measured in $\epem \to \tau^+\tau^-$ events with one of the $\tau$ leptons decaying to three charged hadrons.
A per-track systematic uncertainty of \SI{0.24}{\percent} is included as a normalisation uncertainty of the templates to account for these differences.
The associated systematic uncertainty on \Rmu is \SI{\systrackeff}{\percent}.

Systematic uncertainties due to simulation mismodelling of the detection efficiency of photons and their energies, detector misalignment, and charged-particle momentum correction are each found to be below \SI{0.01}{\percent}.

\subsection{Trigger}

The template yields in each bin are corrected to account for differences in trigger efficiency between data and simulation.
The average of the correction factors that are applied to the simulated events is \num[round-mode=places,round-precision=4]{1.0122} for the muon channel and \num[round-mode=places,round-precision=4]{1.01252} for the electron channel.
Several uncertainties are associated with this correction, summarised in Table~\ref{tab:trigger_avg_summary}.
They include statistical uncertainties from the measured efficiency in data and simulation, and systematic uncertainties associated with the efficiency estimation method.
For the systematic effects, we consider the potential bias introduced if the CDC triggers used to define the reference sample in which the efficiency is measured are not completely independent from the ECL triggers.
This potential bias is obtained from simulation by comparing the efficiency calculated in the reference sample to the absolute efficiency for the entire sample.
In addition, we consider the observed differences in the correction factors when computed with an alternative reference sample with additional CDC triggers used to obtain it.
The latter is the dominant source of uncertainty.
The uncertainties from the various sources are combined in quadrature for each bin and included in the fit model with 44 nuisance parameters under the assumption of independence across momentum bins and between the electron and muon channels.
This assumption is conservative,
since most systematic effects originate from a common source for both channels, which causes a cancellation of the effect when measuring \Rmu.
In particular, the changes in the correction factors when computed with alternative reference triggers are similar for both channels, which indicates a high correlation.
However, some independent effects can still arise due to differences in lepton flavour on the signal hemisphere.
This motivates the more conservative choice of independent uncertainties.
The systematic uncertainty propagating to \Rmu is determined to be \SI{\systrigger}{\percent}.

\begin{table}[ht]
 \renewcommand{\arraystretch}{1.05}
 \centering
 \begin{tabular}{l c @{~}c c @{~}c}
  \hline
                                 & \multicolumn{2}{c}{$\mu$ channel}          & \multicolumn{2}{c}{$e$ channel}                                                              \\
                                 & Range [\%]                                 & Average [\%]                    & Range [\%]                                  & Average [\%] \\
  \hline
  Sample size                    & \SIrange[range-units=single]{0.04}{0.23}{} & \num{0.05}                      & \SIrange[range-units=single]{0.01}{0.08}{}  & \num{0.02}   \\
  Absolute efficiency            & \SIrange[range-units=single]{0.00}{0.10}{} & \num{0.07}                      & \SIrange[range-units=single]{-0.11}{0.01}{} & \num{-0.01}  \\
  Alternative reference triggers & \SIrange[range-units=single]{0.09}{0.20}{} & \num{0.15}                      & \SIrange[range-units=single]{0.09}{0.15}{}  & \num{0.13}   \\
  \hline
 \end{tabular}
 \caption{Range and average of the uncertainties associated with the trigger correction.
  The first row shows the statistical uncertainties, while the other two rows show the uncertainties attributed to systematic effects, as mentioned in the text.
  The uncertainties are given relative to the correction factors.
  The systematic uncertainties are obtained from differences with respect to the nominal correction, the direction of which is given by the sign.
 }
 \label{tab:trigger_avg_summary}
\end{table}

\subsection{Size of the simulated samples and luminosity}

The templates are allowed to vary due to the limited size of the simulated samples, which is parametrised by one nuisance parameter per bin.
The extent of this variation is mainly affected by the size of the simulated \eetautau sample, which is \SI{1}{\invab}.
The resulting systematic uncertainty on \Rmu is \SI{\sysMCSize}{\percent}.

Since the uncertainty on the luminosity measurement affects all distributions, its impact on \Rmu largely cancels.
It does not entirely vanish since the overall change in normalisation of the templates can also be related to a combination of other systematic effects, which in turn can affect \Rmu.
This is a very small effect and of the order of \SI{\syslumi}{\percent}.

\subsection{Consistency checks}

\begin{figure}
 \centering
 \includegraphics[width=.95\linewidth]{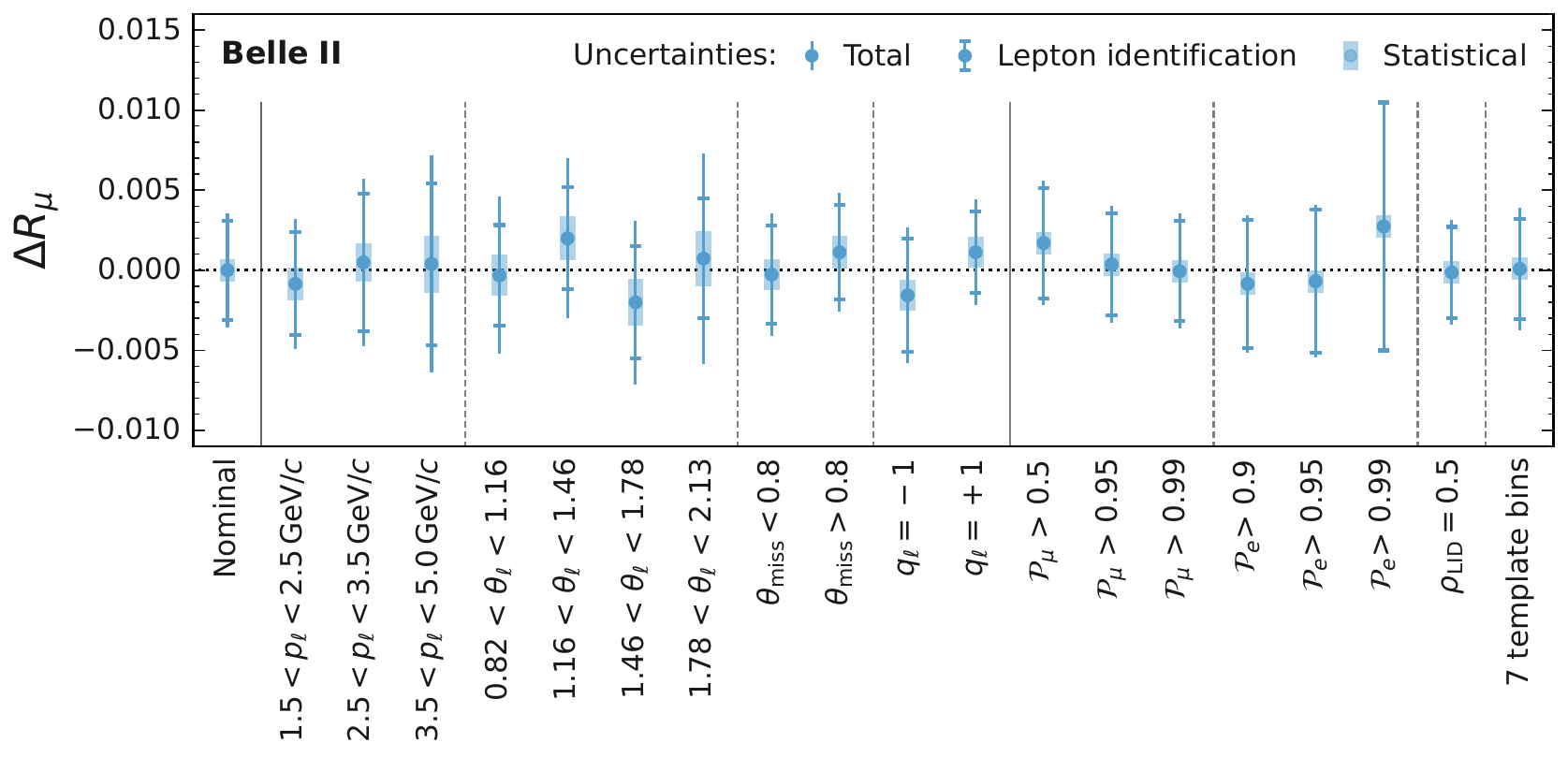}
 \caption{Stability of the fitted value of \Rmu. Shown are deviations $\Delta\Rmu$ from the nominal value for different sub-regions of selected variables, for different requirements on the particle identification discriminators, and for different fit conditions.
  The measurements in each sub-region are statistically independent in the first four groups, separated by dashed vertical lines.
 }
 \label{fig:stability_checks}
\end{figure}

We check the stability of the result throughout various data-taking periods and observe no evidence of time dependence.
We also determine that the value of \Rmu is consistent when varying the event classifier neural network threshold within \num{0.05} around the nominal requirement.
To exclude a potential dependence of the measured \Rmu on the properties of the electrons and muons, we divide the data into sub-regions of various variables. Specifically, we use the lepton charge ($q_{\ell}$), momentum, and polar angle.
Another split of the sample is performed using the polar angle of the missing momentum vector ($\theta_{\text{miss}}$).
To rule out potential dependence on the particle identification requirements, we check the stability of the results using different requirements on the particle identification discriminators $P_e$ and $P_\mu$. Namely, we use $P_e=0.90, 0.95, 0.99$ and $P_\mu=0.50, 0.95, 0.99$ and compare the results to those obtained with the nominal requirements.
We also check the stability of the result when changing the assumed correlation of lepton identification systematic uncertainties ($\rho_{\:\!\text{LID}}$) and the number of bins used to define the templates.
The outcome of some of these checks is shown in Figure~\ref{fig:stability_checks}, where the shaded area indicates the statistical uncertainty, while the error cap marks the systematic uncertainty originating only from lepton identification, and the error bars show the total uncertainties.
We obtain consistent results from all performed checks, indicating no significant unaccounted-for systematic effects.

\section{Results}
\label{sec:results}

\begin{figure}
 \centering
 \includegraphics[width=0.48\linewidth]{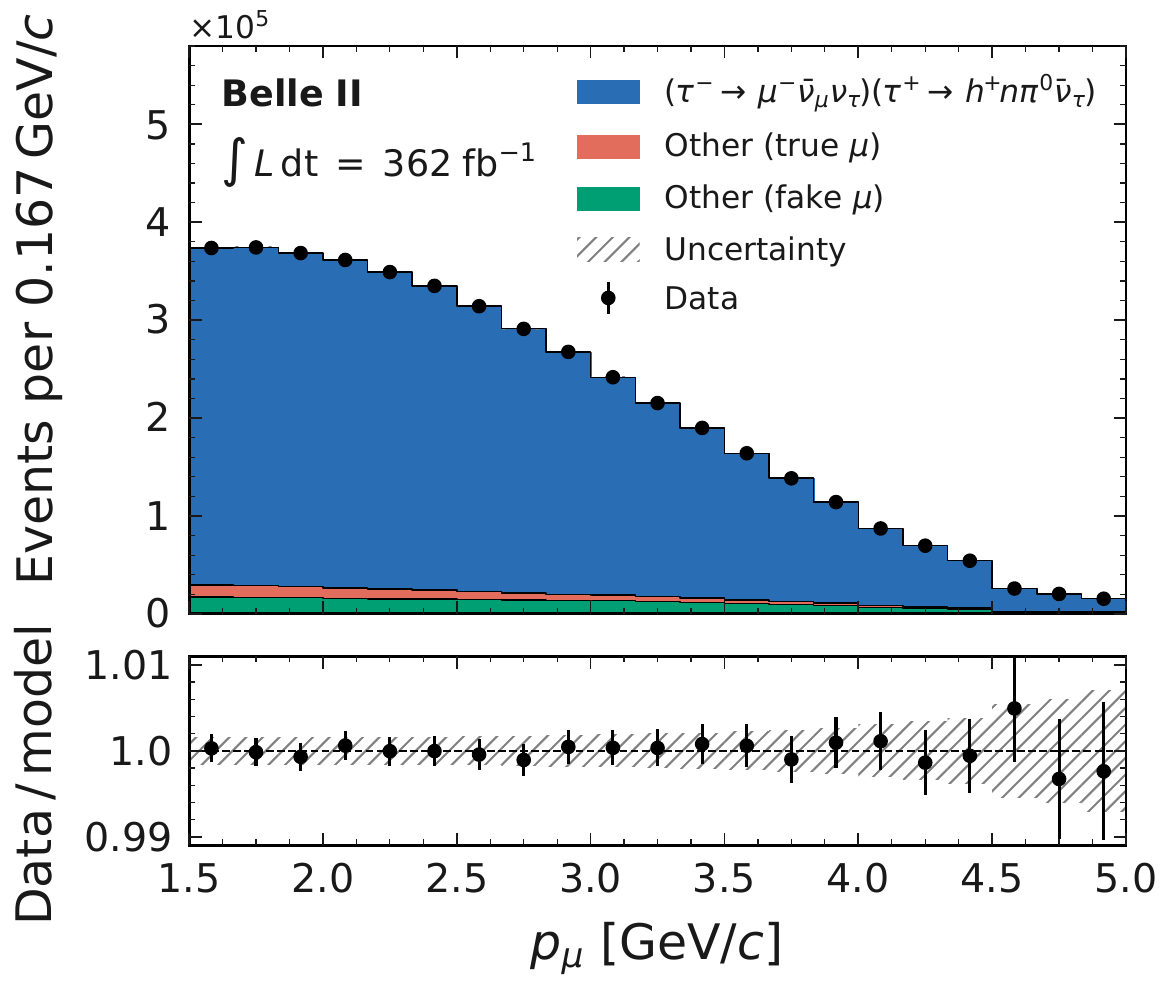}\hfill
 \includegraphics[width=0.48\linewidth]{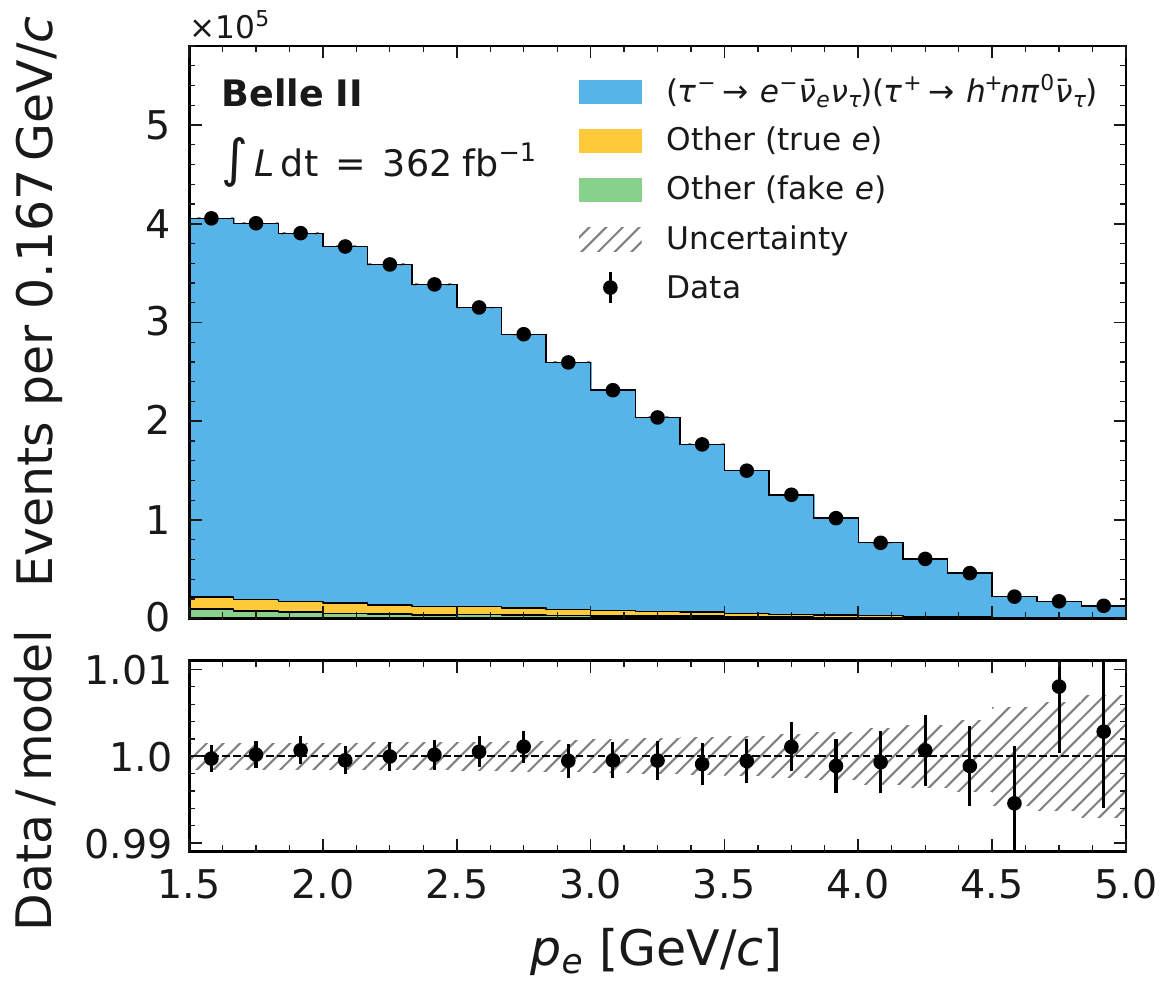}
 \caption{
  Observed momentum distribution for muon (left) and electron (right) candidates with fit results overlaid.
  The lower panel shows the ratio between data and fit results.
  The hatched area indicates the possible variation of the fitted yields due to systematic effects, with the constraints of the nuisance parameters reduced to their fit uncertainties and correlations taken into account.}
 \label{fig:post_fit}
\end{figure}

\begin{figure}
 \centering
 \includegraphics[width=.48\linewidth]{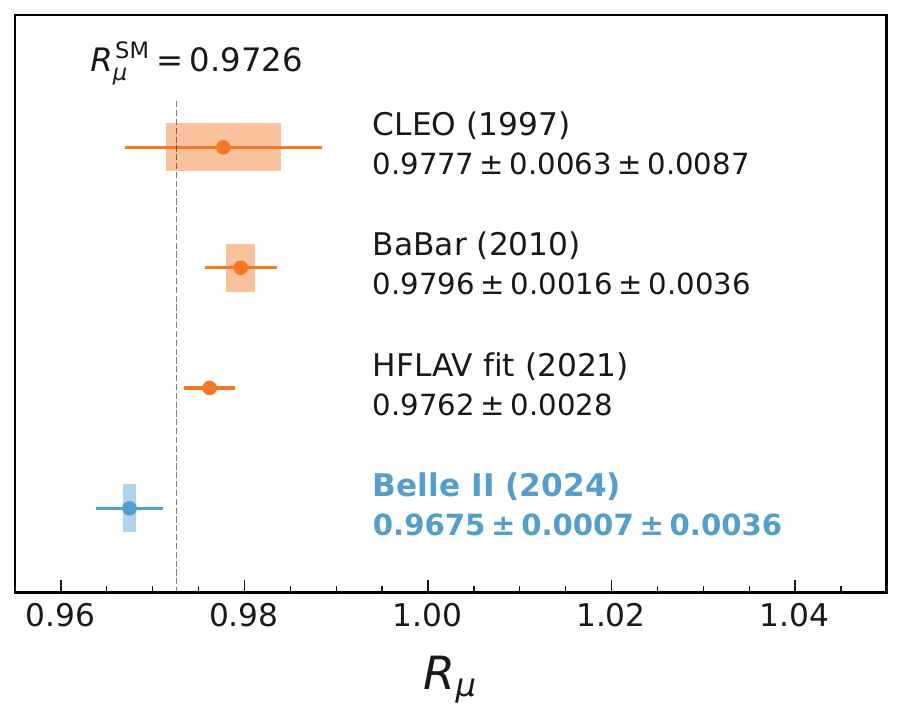}\hfill
 \includegraphics[width=.48\linewidth]{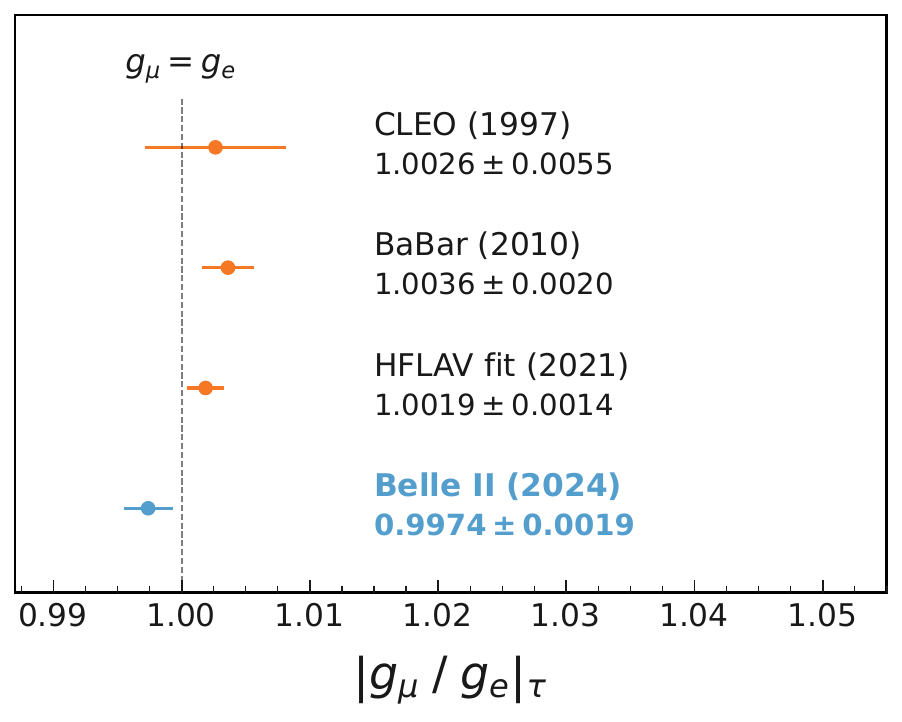}
 \caption{Determinations of \Rmu (left) and \gmuge (right) from previous individual measurements~\cite{CLEO:1996oro, BaBar:2009lyd} and the fit from the Heavy Flavor Averaging Group~\cite{HeavyFlavorAveragingGroup:2022wzx}, compared with the result of this work. The shaded areas represent the statistical uncertainties, while the error bars indicate the total uncertainties. The vertical dashed line indicates the SM prediction, including mass effects.
 }
 \label{fig:worldData}
\end{figure}

The distribution of the muon and electron candidate momentum with fit results overlaid is shown in Figure~\ref{fig:post_fit}.
We measure the ratio of branching fractions
\begin{equation}
 \RmuFull = \result.
\end{equation}
The first uncertainty is statistical, while the second uncertainty is systematic.
The statistical uncertainty is determined by performing the fit with all nuisance parameters fixed to their best-fit values, and the systematic component is calculated by subtraction in quadrature from the total uncertainty obtained with the nominal fit.
This is consistent with the systematic uncertainty obtained in Table~\ref{table:systsum}.
The dominant source of systematic uncertainty is lepton identification.
The measured normalisation yields are \num{4156500 \pm 99650} for \taue and \num{4000190 \pm 99260} for \taumu, with a correlation of \num{0.988}.
The dominant source of uncertainty in the yield estimates is related to the $\pi^0$ reconstruction efficiency, which only affects the tag side.
The value of \Rmu obtained is shown in the left panel of Figure~\ref{fig:worldData} and compared to previous measurements performed by
CLEO~\cite{CLEO:1996oro} and BaBar~\cite{BaBar:2009lyd} as well as the global determination from a fit to all $\tau$ branching fractions~\cite{HeavyFlavorAveragingGroup:2022wzx}.
Our result is consistent with previous measurements and is the most precise measurement from a single experiment to date.
Using Equation~\ref{eq:gmuge}, we translate the measured \Rmu value into the most stringent test of LFU in $\tau$-lepton decays from a single experiment, obtaining $\gmuge=0.9974 \pm 0.0019$ (see the right panel of Figure~\ref{fig:worldData}).

\section{Summary}
\label{sec:summary}

We report a test of light-lepton universality in leptonic $\tau$ decays using a \SI{\lumi}{\invfb} sample of data collected by the Belle II detector at the SuperKEKB \epem collider at a centre-of-mass energy of \SI{\sqrts}{\gev}.
Our result is currently the world's most precise test of light-lepton universality in $\tau$ decays performed by a single experiment and is consistent with the SM.


\acknowledgments
\begin{sloppypar}
This work, based on data collected using the Belle II detector, which was built and commissioned prior to March 2019,
was supported by
Higher Education and Science Committee of the Republic of Armenia Grant No.~23LCG-1C011;
Australian Research Council and Research Grants
No.~DP200101792, 
No.~DP210101900, 
No.~DP210102831, 
No.~DE220100462, 
No.~LE210100098, 
and
No.~LE230100085; 
Austrian Federal Ministry of Education, Science and Research,
Austrian Science Fund
No.~P~34529,
No.~J~4731,
No.~J~4625,
and
No.~M~3153,
and
Horizon 2020 ERC Starting Grant No.~947006 ``InterLeptons'';
Natural Sciences and Engineering Research Council of Canada, Compute Canada and CANARIE;
National Key R\&D Program of China under Contract No.~2022YFA1601903,
National Natural Science Foundation of China and Research Grants
No.~11575017,
No.~11761141009,
No.~11705209,
No.~11975076,
No.~12135005,
No.~12150004,
No.~12161141008,
and
No.~12175041,
and Shandong Provincial Natural Science Foundation Project~ZR2022JQ02;
the Czech Science Foundation Grant No.~22-18469S
and
Charles University Grant Agency project No.~246122;
European Research Council, Seventh Framework PIEF-GA-2013-622527,
Horizon 2020 ERC-Advanced Grants No.~267104 and No.~884719,
Horizon 2020 ERC-Consolidator Grant No.~819127,
Horizon 2020 Marie Sklodowska-Curie Grant Agreement No.~700525 ``NIOBE''
and
No.~101026516,
and
Horizon 2020 Marie Sklodowska-Curie RISE project JENNIFER2 Grant Agreement No.~822070 (European grants);
L'Institut National de Physique Nucl\'{e}aire et de Physique des Particules (IN2P3) du CNRS
and
L'Agence Nationale de la Recherche (ANR) under grant ANR-21-CE31-0009 (France);
BMBF, DFG, HGF, MPG, and AvH Foundation (Germany);
Department of Atomic Energy under Project Identification No.~RTI 4002,
Department of Science and Technology,
and
UPES SEED funding programs
No.~UPES/R\&D-SEED-INFRA/17052023/01 and
No.~UPES/R\&D-SOE/20062022/06 (India);
Israel Science Foundation Grant No.~2476/17,
U.S.-Israel Binational Science Foundation Grant No.~2016113, and
Israel Ministry of Science Grant No.~3-16543;
Istituto Nazionale di Fisica Nucleare and the Research Grants BELLE2;
Japan Society for the Promotion of Science, Grant-in-Aid for Scientific Research Grants
No.~16H03968,
No.~16H03993,
No.~16H06492,
No.~16K05323,
No.~17H01133,
No.~17H05405,
No.~18K03621,
No.~18H03710,
No.~18H05226,
No.~19H00682, 
No.~20H05850,
No.~20H05858,
No.~22H00144,
No.~22K14056,
No.~22K21347,
No.~23H05433,
No.~26220706,
and
No.~26400255,
and
the Ministry of Education, Culture, Sports, Science, and Technology (MEXT) of Japan;
National Research Foundation (NRF) of Korea Grants
No.~2016R1\-D1A1B\-02012900,
No.~2018R1\-A2B\-3003643,
No.~2018R1\-A6A1A\-06024970,
No.~2019R1\-I1A3A\-01058933,
No.~2021R1\-A6A1A\-03043957,
No.~2021R1\-F1A\-1060423,
No.~2021R1\-F1A\-1064008,
No.~2022R1\-A2C\-1003993,
and
No.~RS-2022-00197659,
Radiation Science Research Institute,
Foreign Large-Size Research Facility Application Supporting project,
the Global Science Experimental Data Hub Center of the Korea Institute of Science and Technology Information
and
KREONET/GLORIAD;
Universiti Malaya RU grant, Akademi Sains Malaysia, and Ministry of Education Malaysia;
Frontiers of Science Program Contracts
No.~FOINS-296,
No.~CB-221329,
No.~CB-236394,
No.~CB-254409,
and
No.~CB-180023, and SEP-CINVESTAV Research Grant No.~237 (Mexico);
the Polish Ministry of Science and Higher Education and the National Science Center;
the Ministry of Science and Higher Education of the Russian Federation
and
the HSE University Basic Research Program, Moscow;
University of Tabuk Research Grants
No.~S-0256-1438 and No.~S-0280-1439 (Saudi Arabia);
Slovenian Research Agency and Research Grants
No.~J1-9124
and
No.~P1-0135;
Agencia Estatal de Investigacion, Spain
Grant No.~RYC2020-029875-I
and
Generalitat Valenciana, Spain
Grant No.~CIDEGENT/2018/020;
National Science and Technology Council,
and
Ministry of Education (Taiwan);
Thailand Center of Excellence in Physics;
TUBITAK ULAKBIM (Turkey);
National Research Foundation of Ukraine, Project No.~2020.02/0257,
and
Ministry of Education and Science of Ukraine;
the U.S. National Science Foundation and Research Grants
No.~PHY-1913789 
and
No.~PHY-2111604, 
and the U.S. Department of Energy and Research Awards
No.~DE-AC06-76RLO1830, 
No.~DE-SC0007983, 
No.~DE-SC0009824, 
No.~DE-SC0009973, 
No.~DE-SC0010007, 
No.~DE-SC0010073, 
No.~DE-SC0010118, 
No.~DE-SC0010504, 
No.~DE-SC0011784, 
No.~DE-SC0012704, 
No.~DE-SC0019230, 
No.~DE-SC0021274, 
No.~DE-SC0021616, 
No.~DE-SC0022350, 
No.~DE-SC0023470; 
and
the Vietnam Academy of Science and Technology (VAST) under Grants
No.~NVCC.05.12/22-23
and
No.~DL0000.02/24-25.

These acknowledgements are not to be interpreted as an endorsement of any statement made
by any of our institutes, funding agencies, governments, or their representatives.

We thank the SuperKEKB team for delivering high-luminosity collisions;
the KEK cryogenics group for the efficient operation of the detector solenoid magnet;
the KEK Computer Research Center for on-site computing support; the NII for SINET6 network support;
and the raw-data centers hosted by BNL, DESY, GridKa, IN2P3, INFN,
and the University of Victoria.

\end{sloppypar}

\bibliographystyle{JHEP}
\bibliography{references}

\end{document}